\documentclass[12pt]{article}

\begin{document}

\begin{center}
{\Large\bf{}The gravitational energy-momentum for the super-energy Bel-Robinson tensor}\\
\end{center}

\begin{center}
Lau Loi So
\end{center}

\begin{abstract}
Describing the gravitational energy-momentum, the super-energy Bel-Robinson tensor is the best candidate.  In the past, people seems only explore the lowest order: the electric part $E_{ab}$ and magnetic part $B_{ab}$ for the Riemann tensor.  These two components are related with the static case, however, for the energy transfer situation, one may need to consider the time varying $\dot{E}_{ab}$ and $\dot{B}_{ab}$.  Here we use $(\dot{E}_{ab},\dot{B}_{ab}$) to study the energy-momentum for the Bel-Robinson tensor in a small sphere limit.  Meanwhile, our result illustrates how the gravitational field carries the 4-momentum including this extra information. 
\end{abstract}

\section{Introduction}
The super-energy Bel-Robinson tensor is the most elegant  candidate to describe the gravitational 
energy-momentum in general relativity~\cite{Senovilla}. The well known physical and mathematical properties are: energy positivity, completely
symmetric, trace free and divergence free. Bel and Robinson  demonstrated the gravitational energy-momentum tensor since 1958~\cite{Bel19581st,Bel19582nd,Robinson19581st,Robinson19972nd}. In the beginning, describing the gravitational energy using the Bel-Robinson tensor had the dimensional embarrassment, but this difficulty can be solved by the technique of a small sphere limit~\cite{Horowitz,Szabados}.

Regarding the energy-momentum description of the Bel-Robinson tensor, people seems only examine the lowest order: the electric part $E_{ab}$ and magnetic part $B_{ab}$ for the Riemann curvature tensor.  These two components are related with the static case.  However, if we consider the energy transferring phenomenon such as the tidal heating for the Juipter-Io system or black hole~\cite{Thorne,Zhang,Poisson}, one may need to consider the extra piece of information, the time varying components $\dot{E}_{ab}$ and $\dot{B}_{ab}$.  Here we use $(\dot{E}_{ab},\dot{B}_{ab})$ to study the energy-momentum for the Bel-Robinson tensor, under the limit of a small sphere region.  In addition, our result elucidates how the gravitational field carries the 4-momentum using this extra piece of information.

\section{Technical background}

Throughout this work, we use the same spacetime signature and
notation in MTW~\cite{MTW}, including the geometrical units
$G=c=1$, where $G$ and $c$ are the Newtonian constant and the
speed of light. The Greek letters denote the spacetime and Latin
letters refer to spatial. In vacuum, the Bel-Robinson tensor can
be defined as follows
\begin{eqnarray}
B_{\alpha\beta\mu\nu}:=R_{\alpha\lambda\mu\sigma}R_{\beta}{}^{\lambda}{}_{\nu}{}^{\sigma}
+R_{\alpha\lambda\nu\sigma}R_{\beta}{}^{\lambda}{}_{\mu}{}^{\sigma}
-\frac{1}{8}g_{\alpha\beta}g_{\mu\nu}R^{2}_{\lambda\sigma\rho\tau},\label{27aDec2023}
\end{eqnarray}
where
$R^{2}_{\lambda\sigma\rho\tau}=R_{\lambda\sigma\rho\tau}R^{\lambda\sigma\rho\tau}$.
In order to extract the energy-momentum, one can
use the analog of the ``electric" $E_{ab}$ and ``magnetic"
$B_{ab}$ parts of the Weyl tensor \cite{Carmeli},
\begin{eqnarray}
E_{ab}:=C_{0a0b}, \quad B_{ab}:=\ast{C_{0a0b}},
\end{eqnarray}
where $C_{\alpha\beta\mu\nu}$ is the Weyl conformal tensor and
$\ast{C_{\alpha\beta\mu\nu}}$ is its dual,
\begin{equation}
\ast{C_{\alpha\beta\mu\nu}}=\frac{1}{2}
\epsilon_{\alpha\beta\lambda\sigma} C^{\lambda\sigma}{}_{\mu\nu}.
\end{equation}
(Here
$\epsilon_{\alpha\beta\mu\nu}=\epsilon_{[\alpha\beta\mu\nu]}$ with
$\epsilon_{0123}=\sqrt{-g}$ is the totally anti-symmetric
Levi-Civita tensor, see \cite{MTW}, in particular Eq. 8.10 and Ex.
8.3.) In vacuum using the Riemann tensor
\begin{equation}
E_{ab}=R_{0a0b},\quad B_{ab}=*R_{0a0b}.
\end{equation}
Certain commonly occurring quadratic combinations of the Riemann
tensor components in terms of the electric $E_{ab}$ and magnetic
$B_{ab}$ parts in vacuum are
\begin{equation}
R_{0a0b}R_{0}{}^{a}{}_{0}{}^{b}=E^{2}_{ab},\quad
R_{0abc}R_{0}{}^{abc}=2B^{2}_{ab},\quad
R_{abcd}R^{abcd}=4E^{2}_{ab},\label{27uDec2023}
\end{equation}
where $E^{2}_{ab}:=E_{ab}E^{ab}$ and similarly for $B^{2}_{ab}$. In particular, the Riemann squared tensor can then be written
\begin{equation}
R^{2}_{\lambda\sigma\rho\tau} =8(E^{2}_{ab}-B^{2}_{ab}).\label{27vDec2023}
\end{equation}

Within a weak field the metric tensor can be decomposed as
$g_{\alpha\beta}=\eta_{\alpha\beta}+h_{\alpha\beta}$, and its
inverse is $g^{\alpha\beta}=\eta^{\alpha\beta}-h^{\alpha\beta}$.
Here we only consider the lowest order (see (8) at~\cite{Zhang}),
the metric components can be written as
\begin{eqnarray}
h^{00}&=&\frac{3}{r^{5}}I_{ab}x^{a}x^{b}-E_{ab}x^{a}x^{b},\label{27xDec2023}\\
h^{0j}&=&\frac{4}{r^{5}}\,\epsilon^{j}{}_{pq}J^{p}{}_{l}\,x^{q}x^{l}
+\frac{2}{3}\epsilon^{j}{}_{pq}B^{p}{}_{l}x^{q}x^{l}
+\frac{2}{r^{3}}\dot{I}^{j}{}_{a}\,x^{a}
+\frac{10}{21}\dot{E}_{ab}x^{a}x^{b}x^{j}
-\frac{4}{21}\dot{E}^{j}{}_{a}x^{a}r^{2},\label{27yDec2023}\\
h^{ij}&=&\eta^{ij}h^{00}+\bar{h}^{ij},\label{27zDec2023}
\end{eqnarray}
where
\begin{eqnarray}
\bar{h}^{ij}=\frac{8}{3r^{3}}\epsilon_{pq}{}^{(i}\dot{J}^{j)p}x^{q}
+\frac{5}{21}x^{(i}\epsilon^{j)}{}_{pq}\dot{B}^{q}{}_{l}x^{p}x^{l}
-\frac{1}{21}r^{2}\epsilon_{pq}{}^{(i}\dot{B}^{j)\,q}x^{p}.
\end{eqnarray}
Zhang used $\bar{h}^{\alpha\beta}$ for the manipulation while we
prefer using $h^{\alpha\beta}$, the transformation is as follows
\begin{eqnarray}
\bar{h}^{\alpha\beta}=h^{\alpha\beta}-\frac{1}{2}\eta^{\alpha\beta}h.
\end{eqnarray}
The corresponding first order harmonic gauge is
$\partial_{\beta}\bar{h}^{\alpha\beta}=0$. Moreover, we will
substitute the mass quadrupole moment $I_{ij}$ and current
quadrupole moment $J_{ij}$ as determined by
Poisson~\cite{Poisson}:
\begin{eqnarray}
I^{ij}=\frac{32}{45}M^{6}\dot{E}^{ij},\quad{}
J^{ij}=\frac{8}{15}M^{6}\dot{B}^{ij},
\end{eqnarray}
where $M$ is the mass of the black hole. The value of the tidal
heating is something like
$I^{ij}\dot{E}_{ij}+J^{ij}\dot{B}_{ij}\simeq{}\dot{E}^{2}_{ij}+\dot{B}^{2}_{ij}$,
where $\dot{E}^{2}_{ij}$ means $\dot{E}^{ij}\dot{E}_{ij}$,
likewise for $\dot{B}^{2}_{ij}$. As the tidal heating is a kind of energy dissipation, obviously that both
$\dot{E}^{2}_{ab}$ and $\dot{B}^{2}_{ab}$ 
indicated the physical interaction for the energy transfer.

\section{The energy-momentum for the Bel-Robinson tensor}

Referring to (\ref{27aDec2023}), the usual gravitational energy density presentation for the Bel-Robinson tensor is
\begin{eqnarray}
B_{0000}&=&2R_{0a0b}R_{0}{}^{a}{}_{0}{}^{b}-\frac{1}{8}g_{00}g_{00}R^{2}_{\lambda\sigma\rho\tau}\nonumber\\
&=&E^{2}_{ab}+B^{2}_{ab}.
\end{eqnarray}
This basic result can be obtained from (\ref{27uDec2023}) to 
(\ref{27vDec2023}) 
or from (\ref{27xDec2023}) to
(\ref{27zDec2023}).
As mentioned before, this presentation had the dimensional problem, the appropriate method is the small sphere limit. Here we use this small sphere technique, including $(\dot{E}_{ab},\dot{B}_{ab})$ and ignored the higher order, the energy is
\begin{eqnarray}
E&=&\frac{1}{16\pi}\int_{V}B_{00mn}\,x^{m}x^{n}\,d^{3}x
\nonumber\\
&=&\frac{1}{16\pi}\int_{V}\left[
2(R_{0amb}R_{0}{}^{a}{}_{n}{}^{b}
-R_{0a0m}R_{0}{}^{a}{}_{0n})
-\frac{1}{8}g_{00}g_{mn}R^{2}_{\lambda\sigma\rho\tau}
\right]x^{m}x^{n}\,d^{3}x\nonumber\\
&=&\frac{1}{16\pi}\int_{V}\left[
\begin{array}{ccc}
(E^{2}_{ab}+B^{2}_{ab})\,r^{2}
-2(E^{c}{}_{c\,q}E_{c\,q}+B^{c}{}_{p}B_{c\,q})\,x^{p}x^{q}\quad\\
+\frac{2}{9}(
\dot{E}_{c\,d}\dot{E}_{p\,q}\,x^{c}x^{d}x^{p}x^{q}-4\dot{E}^{c}{}_{p}\dot{E}_{c\,q}\,x^{p}x^{q}\,r^{2}+2\dot{E}^{2}_{ab}\,r^{4})\\
+\frac{2}{9}
(\dot{B}_{c\,d}\dot{B}_{p\,q}\,x^{c}x^{d}x^{p}x^{q}-4\dot{B}^{c}{}_{p}\dot{B}_{c\,q}\,x^{p}x^{q}\,r^{2}+2\dot{B}^{2}_{ab}\,r^{4})\\
\end{array}
\right]d^{3}x\nonumber\\
&=&\frac{1}{60}(E^{2}_{ab}+B^{2}_{ab})\,r^{5}+\frac{2}{315}(\dot{E}^{2}_{ab}+\dot{B}^{2}_{ab})\,r^{7}.
\end{eqnarray}
Note that the sign of this energy guarantees the positivity.   
Moreover, that extra piece quantity $(\dot{E}^{2}_{ab}+\dot{B}^{2}_{ab})\,r^{7}$ illustrates how the gravitational field carries the energy.

Likewise, we demonstrate the momentum for this Bel-Robinson tensor.  Recall the usual presentation for the momentum density
\begin{eqnarray}
B_{000i}=2R_{0a0b}R_{0}{}^{a}{}_{i}{}^{b}=2\epsilon_{iab}E^{ac}B^{b}{}_{c}.
\end{eqnarray}
Again, this result can be computed from (\ref{27uDec2023}) to 
(\ref{27vDec2023}) 
or from (\ref{27xDec2023}) to
(\ref{27zDec2023}). Note that the dimensional problem still exist. Using the method of a small sphere limit, including $(\dot{E}_{ab},\dot{B}_{ab})$ and omitted the higher order, the momentum is
\begin{eqnarray}
P_{i}&=&\frac{1}{16\pi}\int_{V}B_{0imn}\,x^{m}x^{n}d^{3}x
\nonumber\\
&=&\frac{1}{16\pi}\int_{V}
2(R_{0amb}R_{i}{}^{a}{}_{n}{}^{b}-R_{0a0m}R_{i}{}^{a}{}_{0n})\,x^{m}x^{n}d^{3}x\nonumber\\
&=&\frac{1}{16\pi}\int_{V}\left[
\begin{array}{ccc}
2\epsilon_{c\,p\,q}(E^{c}{}_{i}B^{p}{}_{d}\,x^{q}x^{d}
-E_{c\,d}B^{p\,d}\,x^{q}x_{i}
+\frac{2}{3}E^{c}{}_{d}B^{p}{}_{i}\,x^{q}x^{d})\\
+\frac{2}{3}\epsilon_{i\,p\,q}(E_{c\,d}B^{p\,c}\,x^{q}x^{d}
+E^{p}{}_{c}B^{q}{}_{d}\,x^{c}x^{d})\quad\quad\quad\quad\quad\quad\\
\end{array}
\right]d^{3}x\nonumber\\
&&+\frac{1}{16\pi}\int_{V}\frac{1}{9}\left[
\begin{array}{ccc}
4\epsilon_{c\,p\,q}\dot{E}^{c}{}_{d}\dot{B}^{p}{}_{l}\,x^{q}x^{l}x^{d}x_{i}
\quad\quad\quad\quad\quad\quad
\quad\quad\quad\quad\quad\quad\quad\\
+\epsilon_{c\,p\,q}(\dot{E}^{c}{}_{i}\dot{B}^{p}{}_{d}\,x^{q}x^{d}
+2\dot{E}^{c\,d}\dot{B}^{p}{}_{i}\,x^{q}x^{d}
-5\dot{E}^{c\,d}\dot{B}^{p}{}_{d}\,x^{q}x_{i})\,r^{2}\\
+\epsilon_{i\,p\,q}(
3\dot{E}^{p}{}_{c}\dot{B}^{q}{}_{d}\,x^{c}x^{d}
-\dot{E}_{c\,d}\dot{B}^{p}{}_{c}\,x^{q}x^{d}
-3\dot{E}^{p\,c}\dot{B}^{q}{}_{c}\,r^{2})\,r^{2}\quad
\end{array}
\right]d^{3}x\nonumber\\
&=&-\left[\frac{1}{30}\epsilon_{iab}E^{ac}B^{b}{}_{c}\,r^{5}
+\frac{4-\delta}{315}\epsilon_{iab}\dot{E}^{ac}\dot{B}^{b}{}_{c}\,r^{7}\right],
\end{eqnarray}
where $\delta=\frac{1}{6}$.  In addition, that extra piece quantity 
$\epsilon_{iab}\dot{E}^{ac}\dot{B}^{b}{}_{c}\,r^{7}$ explains how the gravitational field carries the momentum.

After obtained the extra piece information of the energy-momentum for the Bel-Robinson tensor, we emphasis that it satisfies the future pointing and non-spacelike property.

\section{Conclusion}

It is known that the super-energy Bel-Robinson tensor is the best candidate to describe the gravitational energy-momentum in general relativity. According to the phenomenon of the tidal heating, the Jupiter-Io system or the black hole, we observed that the gravitational energy should be related with the time varying components of the electric part $\dot{E}_{ab}$ and magnetic part $\dot{B}_{ab}$ for the Riemann curvature tensor. Here we used the Bel-Robinson tensor to demonstrate the gravitational energy-momentum, including these extra components 
$(\dot{E}_{ab},\dot{B}_{ab})$, 
we find that it gives a desirable result under a small sphere limit, i.e., the 4-momentum are future pointing and non-spacelike.  Moreover, our result also illustrates that how the gravitational field carries the energy-momentum through the empty spacetime.

\end{document}